\documentclass{PoS}\PoS{PoS(LAT2005)077}
\usepackage{epsfig}
\title{Quark masses from quenched overlap fermions\footnote{Preprint DESY
    05-178, Edinburgh 2005/14} } \ShortTitle{Quark masses from quenched overlap fermions }
\author{\speaker{Martin G\"urtler}, Thomas Streuer\\%
  NIC/DESY Zeuthen\\%
  E-mail: \protect\email{\{martin.guertler,thomas.streuer\}@desy.de}}
\author{Gerrit Schierholz\\%
  DESY Hamburg\\%
  E-mail: \email{gerrit.schierholz@desy.de}} 
\author{David
  Galletly, Roger Horsley\\%
  University of Edinburgh\\%
  E-mail: \email{\{galletly,rhorsley\}@ph.ed.ac.uk}} 
\author{Paul E. L. Rakow\\%
  University of Liverpool\\%
  E-mail: \email{rakow@amtp.liv.ac.uk}}
\author{QCDSF collaboration}

\abstract{We compute light and strange
  quark masses for quenched overlap fermions at two values of the gauge
  coupling. The renormalisation is done non-perturbatively. We test the
  predictions of quenched chiral perturbation theory for the quark mass
  dependence of the hadron spectrum and see evidence for the existence of
  chiral logs.}
\FullConference{XXIIIrd International Symposium on Lattice Field Theory\\
  25-30 July 2005\\
  Trinity College, Dublin, Ireland}
\newcommand{\fm}{\ensuremath{\;\mathrm{fm}}}
\newcommand{\mev}{\ensuremath{\;\mathrm{MeV}}}
\newcommand{\gev}{\ensuremath{\;\mathrm{GeV}}}
\newcommand{\msbar}{{\overline{MS}}} 

\begin{document}
\bibliographystyle{JHEP-2}
\section{Introduction}
Overlap fermions provide the opportunity to investigate QCD at small quark
masses. Being computationally extremely expensive, no dynamical results on
physically relevant lattices are available yet. In this work, we present
spectrum and quark mass results of a quenched study. We calculate light and
strange quark masses, and we check several predictions of quenched chiral
perturbation theory. Results for nucleon matrix elements are presented
in~\cite{streuer:_nucleon}.
\section{Action, lattices and the scale}
\begin{figure}[bth]
  \begin{minipage}[t]{0.45\linewidth}
    \epsfig{file=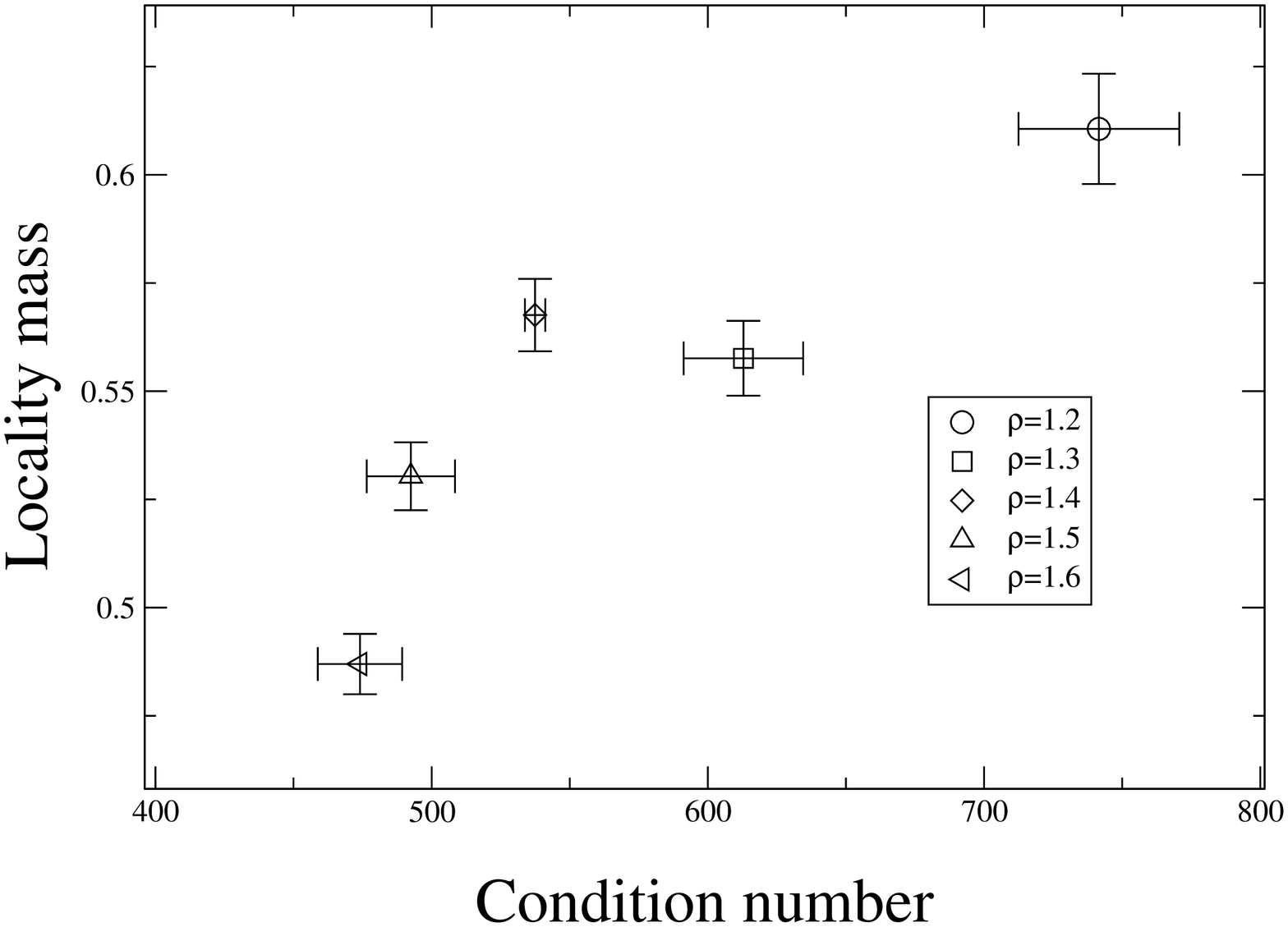, width=5.5cm, angle=270,clip="}
    \caption{
      \label{fig:condloc}
      Condition number and localisation mass (1-norm) for $\beta=8.45$, 10 eigenvalues
      projected out.}
  \end{minipage}
  \hspace{0.05\linewidth}
  \begin{minipage}[t]{0.45\linewidth}
      \epsfig{file=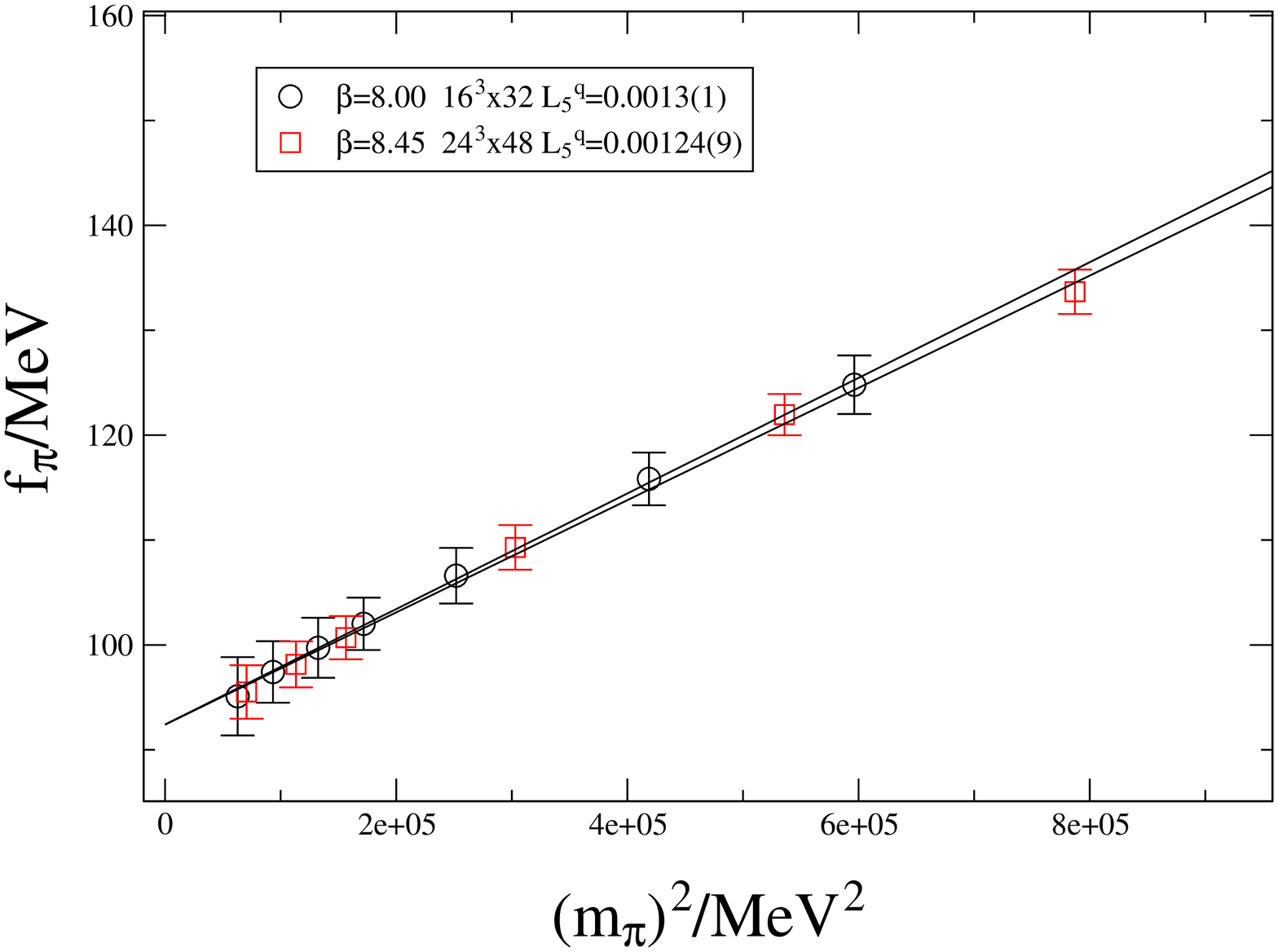,width=5.5cm, angle=270, clip=}
      \caption{
        \label{fig:fpifits}
        Quenched chiral fits to $f_\pi$.}
    \end{minipage}
\end{figure}

Our overlap operator is 
\begin{equation}
D=\left(1-\frac{am_q}{2 \rho} \right) D_N + m_q ,\qquad\qquad D_N=\frac{\rho}{a}\left(1+{X \left(X^\dagger X\right)}^{-1/2}  \right),
\end{equation}
with the Wilson kernel operator ${X}=D_W-\frac{\rho}{a}$. We use a polynomial
approximation (see~\cite{Giusti:2002sm}) for ${X \left(X^\dagger
    X\right)}^{-1/2}$, the degree of which is adjusted properly to get an
overall residual in the inversion of the overlap better than $10^{-7}$ . The
value of $\rho$ is chosen to satisfy the requirements of low condition number
and good localisation properties. A way to improve the condition number
is the exact treatment of the lowest eigenvalues. 
Our choice, $\rho=1.4$, is a compromise between small condition
number and large localisation mass ({\it i.e.} we aim at the upper left hand
corner in Fig.~\ref{fig:condloc}).

We use the 1-loop tadpole improved L\"uscher-Weisz gauge
action~\cite{Luscher:1985xn}. 
  \begin{equation}
    S[U]=\frac{6}{g^2}\left[ c_0 
    \sum_{\rm plaq}\frac{1}{3}\,
    \mbox{Re}\, 
    \mbox{Tr}\, (1-U_{\rm plaq}) 
    + c_1 \sum_{\rm rect}\frac{1}{3}\, \mbox{Re}\, 
    \mbox{Tr}\, (1-U_{\rm rect})  
    + c_2 \sum_{\rm par}\frac{1}{3}\, 
    \mbox{Re}\, \mbox{Tr}\, (1-U_{\rm par}) \right] .
  \end{equation}
The lattice gauge coupling $\beta$ is defined as $\beta=6c_0/g^2$, and the values
of $c_1$ and $c_2$ are fixed
by the value of $c_0$ and the plaquette expectation value $u_0$.
The benefits of the L\"uscher-Weisz compared to the Wilson gauge action are a better
condition number of $X^\dag X$  (as long as no gauge smearing is involved) and absence of dislocations.

We use Jacobi smeared sources ($\kappa=0.21$, $n=50$).
To set the scale we compute
\begin{equation}
  \label{eq:1}
  f_\pi(m_q) = \frac{m_q}{m_\pi^{3/2}}\frac {A_{sl}}{\sqrt{A_{ss}}}  ,
\end{equation}
for each of our quark masses, where $m_\pi$ and $A_{sl,ss}$ are the mass and amplitude from cosh fits to the
smeared-local and smeared-smeared pseudoscalar correlators. We  extrapolate
the lattice numbers to the chiral limit by the leading order formula 
\begin{equation}
  f_\pi(m_q)=f_\pi\;\left(1+4L_5^q
  \frac{m_\pi^2}{f_\pi^2}\right),
\label{eq:6}
\end{equation}
The resulting values of $a$ are given in Table~\ref{tab:lattices} (for the values of $L_5^q$
see the legend of Fig.~\ref{fig:fpifits}). Note, that the lattice constant is
about $10\%$ larger than in~\cite{Gattringer:2001jf} where the Sommer scale
was used. Such a scale ambiguity is a familiar phenomenon of quenched
simulations.

\begin{table}[htb]
  \centering
  \begin{tabular}{c|c|c|c|c}
    $\beta$ & lattice            &  cfg. & $a/\fm$ & $a^{-1}/\mev$   \\
    \hline
    8.0     & $16^3 \times 32$   &  $230$   &$0.153(3)$ & $1290(30)$ \\
    8.45    & $16^3 \times 32$   &  $250$   &$0.105(2)$ & $1870(40)$ \\
    8.45    & $24^3 \times 48$   &  $200$   &$0.105(1)$ & $1870(20)$ \\
  \end{tabular}
\caption{Lattices used in the simulation. The scale was determined via
  $f_\pi$ (see Fig.~\protect\ref{fig:fpifits}).}
\label{tab:lattices}
\end{table}

\section{Quark masses}
\label{sec:quark-masses}
\begin{figure}[tbh]
  \begin{minipage}[t]{0.45\linewidth}
    \epsfig{file=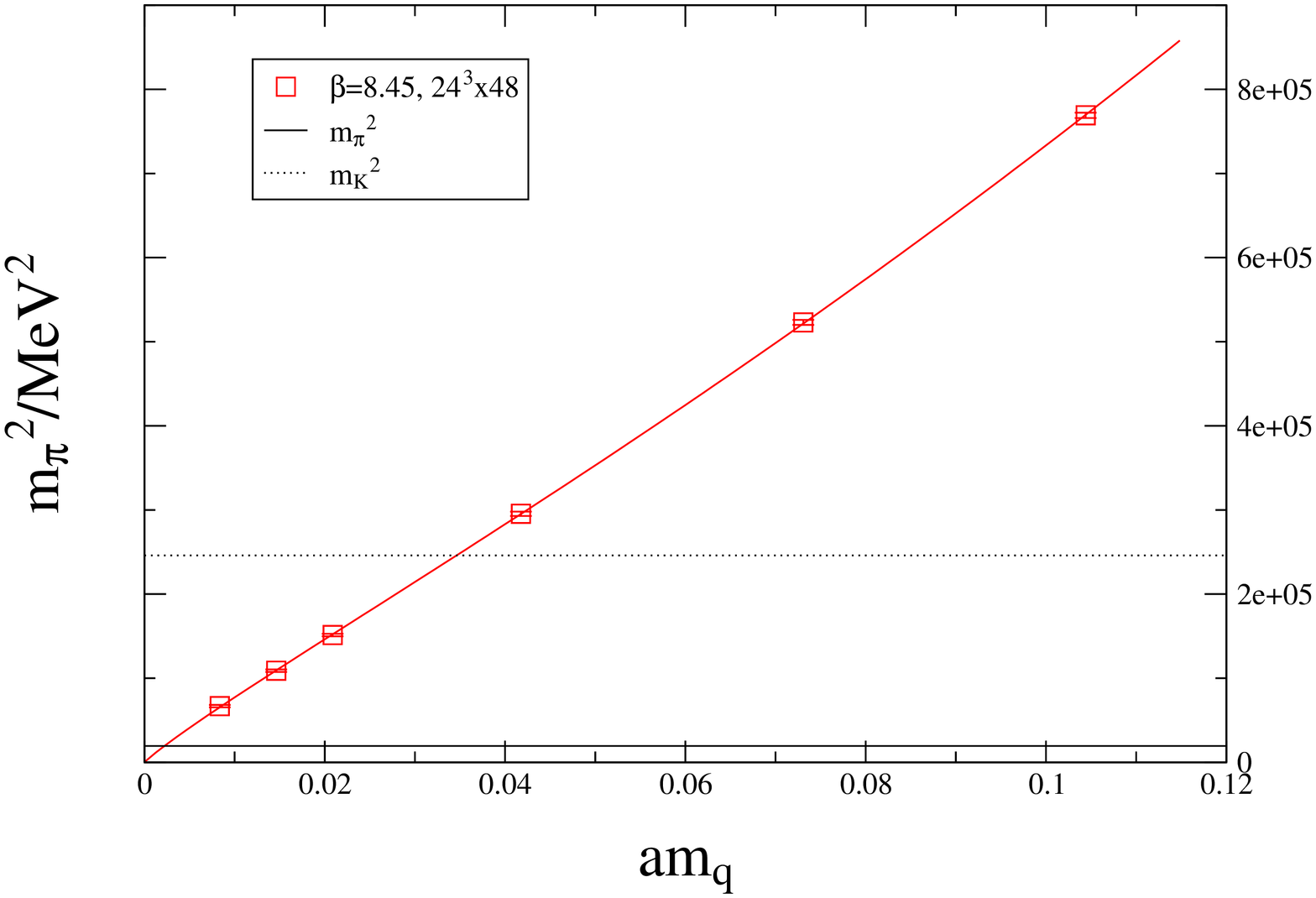,width=5.5cm, angle=270, clip=}
    \caption{
      \label{fig:mpi2}
      The squared pseudoscalar mass as function of $m_q$ together wit the fit Eq.~(\protect\ref{eq:mq}). }
  \end{minipage}
  \hspace{0.05\linewidth}
  \begin{minipage}[t]{0.45\linewidth}
      \epsfig{file=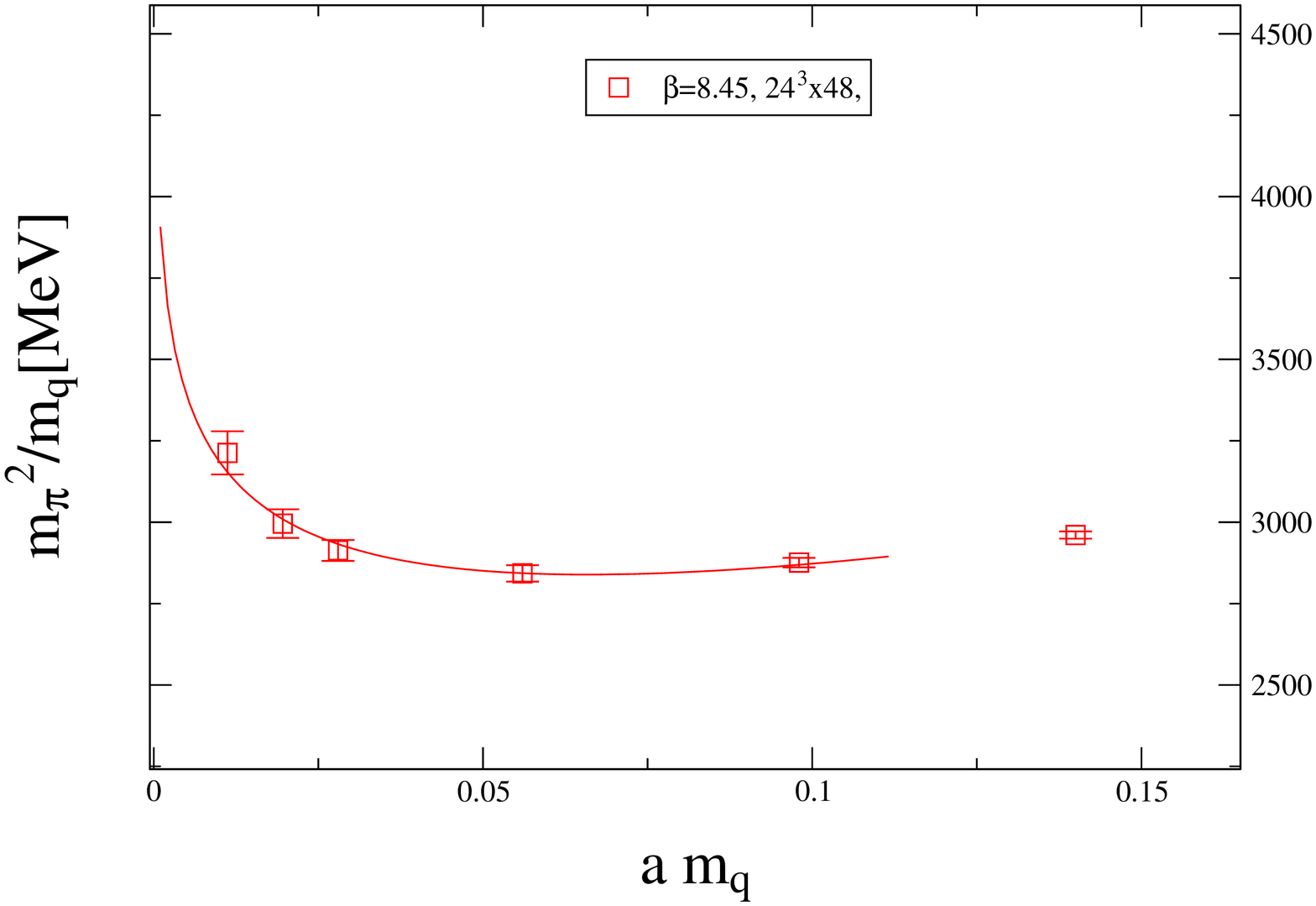,width=5.5cm, angle=270, clip=}
      \caption{
        \label{fig:chirallogs}
        A plot of $m_\pi^2/m_q$ reveals the existence of quenched chiral logs.}
    \end{minipage}
\end{figure}

The best signal for the pseudoscalar mass is obtained from the correlator of
the time component of the axial current. An example of the results is
displayed in Fig.~\ref{fig:mpi2}.
The quenched chiral perturbation theory prediction for $m_\pi$ reads
\begin{equation}
  \label{eq:5}
  m_\pi^2=A m_q
      \left(1-\delta\left( \ln\left(A
            m_q/\Lambda_\chi^2\right)+1\right)\right)+{\cal O}\left(m_q^2\right),
\end{equation}
where the quenched chiral log appears with a prefactor $\delta$ giving rise to
a singularity in $m_\pi^2/m_q$. The fit is shown in  Fig.~\ref{fig:chirallogs}. Fixing
$\Lambda_\chi$ to $1\gev$, we obtain the values of $\delta$ in
Table~\ref{tab:delta}.
\begin{table}[htb]
  \centering
  \begin{tabular}{c|c}
    lattice &  $\delta$ \\
    \hline
    $8.00,\,16^3\times 32$ & $0.16(4)$\\
    $8.45,\,16^3\times 32$ & $0.3(1)$\\
    $8.45,\,24^3\times 48$ & $0.15(3)$\\
  \end{tabular}
  \caption{The chiral log parameter $\delta$.}
  \label{tab:delta}
\end{table}
Although the chiral  logs will strongly affect the determination of
$m_\ell=1/2 (m_u+m_d)$,
we will ignore this problem in the following.

The fit functions we use to determine the bare light and strange quark masses are
\begin{eqnarray}
  m_\pi^2 &=& A m_{\ell}+B m_{\ell} \ln m_{\ell} + C m_{\ell}^2,\nonumber\\
  m_K^2   &=& A \frac{m_\ell + m_s}2
  + B \frac {m_\ell + m_s}2
  \left(\frac{m_s \ln m_s - m_\ell \ln m_\ell}{m_s - m_\ell} -1\right)
  + C \left({\frac{m_\ell + m_s}2}\right)^2.
  \label{eq:mq}
\end{eqnarray}

\begin{figure}[tbh]
\vspace{-7mm}
  \begin{minipage}[t]{0.45\linewidth}
    \epsfig{file=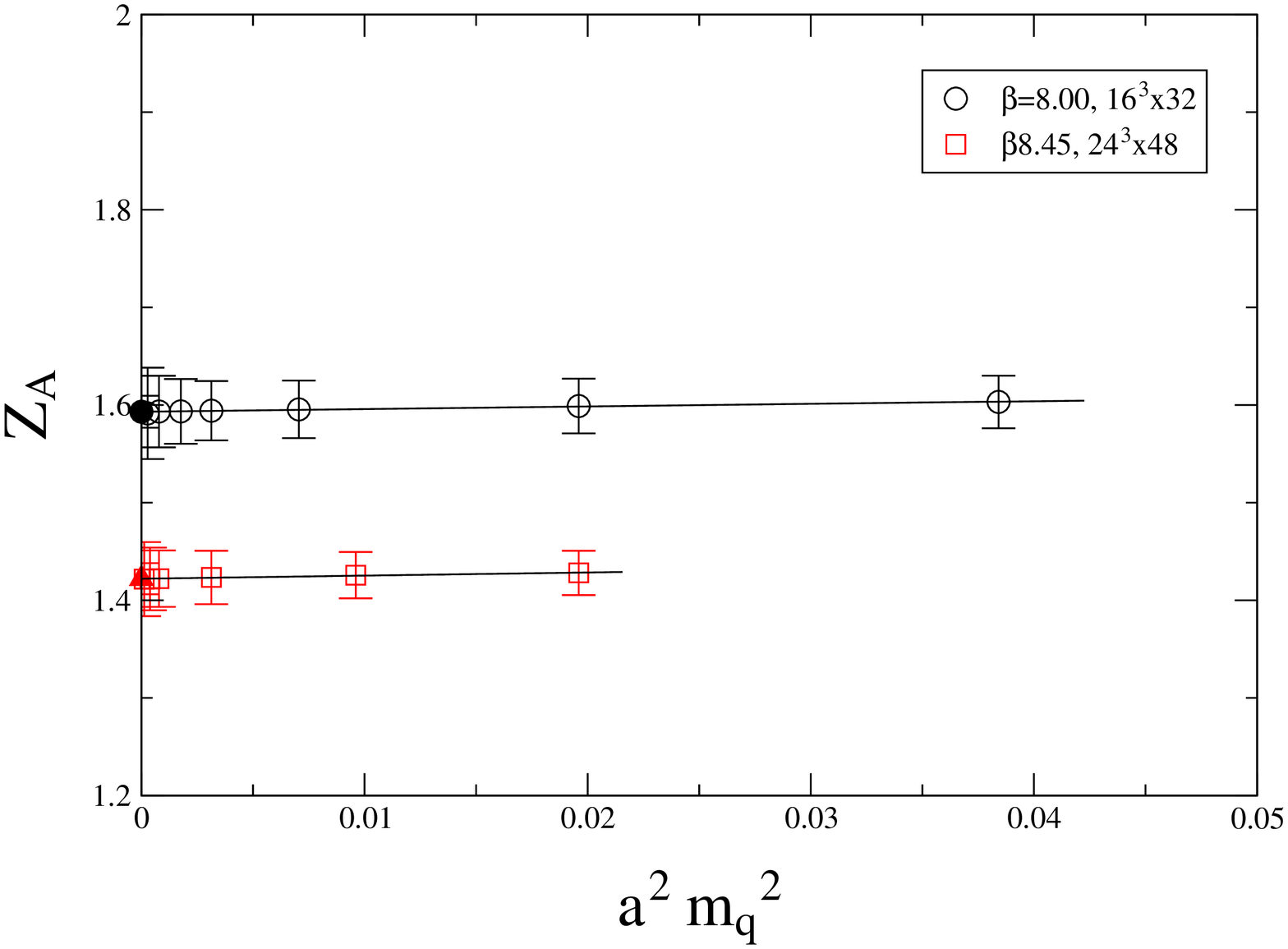,width=5.5cm, angle=270, clip=}
    \caption{
      \label{fig:ZA}
      $Z_A$ from the axial Ward identity.}
  \end{minipage}
\hspace{0.05\linewidth}
  \begin{minipage}[t]{0.45\linewidth}
    \epsfig{file=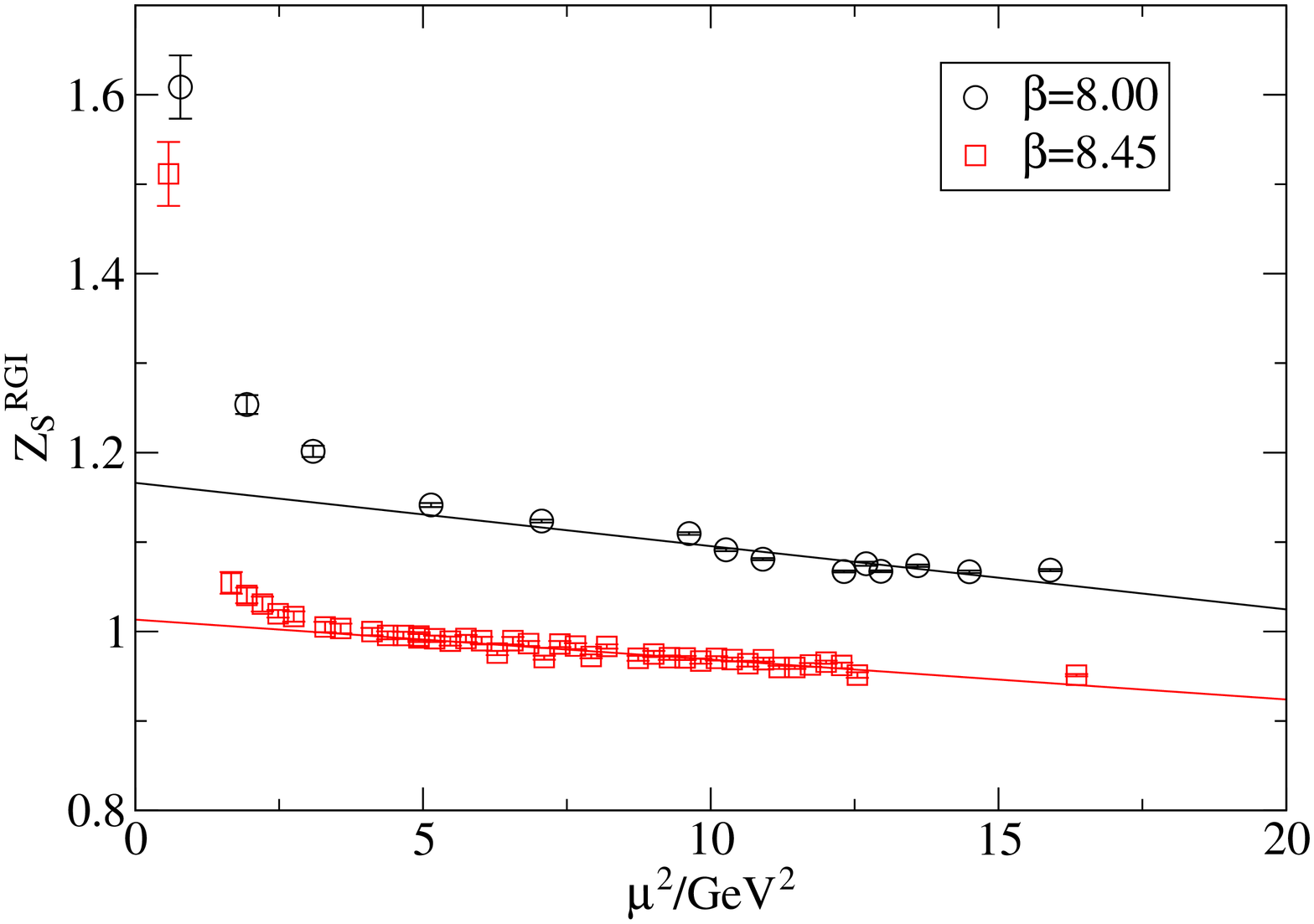,width=5.5cm, angle=270, clip=}
    \caption{    
      \label{fig:ZSRGI}
      $Z_S^\mathrm{RGI}$ for both values of $\beta$}
  \end{minipage}
\end{figure}

The renormalisation is done non-pertubatively\footnote{A perturbative
  calculation of the renormalisation factors can be found
  in~\cite{Horsley:2004mx}.} in the $RI'-MOM$ scheme with a variant of the
method introduced in~\cite{Martinelli:1994ty}, using momentum
sources\cite{Gockeler:1998ye}. This variant requires only small statistics,
and one does not need to perform inversions for each operator under
investigation (although one needs to invert the Dirac operator for each
momentum separately). We implement the renormalisation condition for the
vertex
\begin{equation}
  \label{eq:ren_cond}
    \mbox{Tr}\left(\Gamma_O^{{\mathrm{ren}}}\Gamma_{O,{\mathrm{Born}}}^{-1}\right)=12,
\end{equation}
on the lattice. Putting in the definition of the renormalised vertex we obtain
\begin{equation}
  \label{eq:3}
  \frac1{Z_m}=Z_S=\frac{Z_q}{\frac1{12} \mbox{Tr}\left(\Gamma_S
        \Gamma_{S,{\mathrm{Born}}}^{-1}\right)}.
\end{equation}
In the following, we use $\Lambda_O$ for the denominator of the r.h.s of
Eq.~(\ref{eq:3}) for $O=\{A,V,P,S\}$.
The wave function renormalisation $Z_q$ is computed from $Z_q=Z_A \Lambda_A$,
where we use the axial Ward identity to compute $Z_A$, so that it can be
extrapolated
\begin{equation}
  \label{eq:2}
  Z_A=\lim_{m_q \to 0}\lim_{t \to \infty }\frac
  {2 m_q \left<P\left(t\right)P\left(0\right)\right>}
  {\left<\partial_4 A_4\left(t\right)P\left(0\right)\right>}.
\end{equation}
The chiral extrapolation is illustrated in Fig.~\ref{fig:ZA}. 

We now proceed with the computation of $\Lambda_A $and $\Lambda_S$. We exploit
the identity is $\Lambda_S=\Lambda_P$, a consequence of chiral symmetry. On the
lattice both are modified by zero mode effects $\propto 1/(a m_q)^2$, which
are finite volume artefacts. Additionally, spontaneous chiral symmetry
breaking produces a $1/(a m_q)$ contribution to $\Lambda_P^\mathrm{latt}$.
Thus, we extract $\Lambda_S$ as the common constant from fits
\begin{eqnarray}
  \label{eq:4}
  \Lambda_P^\mathrm{latt}&=& 
  \frac {p_1}{a^2 m_q^2}+\frac {p_2}{a m_q}+\Lambda_S +p_4 a^2 m_q^2,\nonumber\\
  \Lambda_S^\mathrm{latt}&=& 
  \frac {s_1}{a^2 m_q^2}+\Lambda_S +s_3 a^2 m_q^2.
\end{eqnarray}
for each value of $\mu^2$. $\Lambda_A$ is computed in a similar fashion from
linear fits to $\Lambda_A^\mathrm{latt}$ and $\Lambda_V^\mathrm{latt}$ with a
common constant. In the next step we need to identify the region where
non-perturbative effects are small. To this end we remove the renormalisation
group running from the $RI'-MOM$ renormalisation constant $Z_S=Z_A
\Lambda_A/\Lambda_S$ using the 4-loop results from~\cite{Gracey:2003yr}. The
result is shown in Fig.~\ref{fig:ZSRGI}, where the linear regions are clearly
identified. We extract the intercept and quote the results in the $\msbar$
scheme at $\mu=2\gev$ in Table~\ref{tab:ZS}.
\begin{table}[htb]
  \begin{minipage}[t]{0.6\linewidth}
    \begin{tabular}{c|c|c|c}
      $\beta$ & $Z_S^{RGI}$    &  slope  &$Z_S^\msbar(2\gev$)  \\
      \hline
      $8.00$    &  $1.18(2)$  &  $-0.007(2)$ & $0.85(2)$ \\
      $8.45$    &  $1.02(1)$  &  $-0.004(1)$ & $0.73(1)$ \\
    \end{tabular}
    \caption{
      \label{tab:ZS}
      $Z_S$}
  \end{minipage}
  \begin{minipage}[t]{0.4\linewidth}
    \begin{tabular}{c|c|c}
      $\beta$ & $m_\ell^{\msbar}$    &  $m_s^{\msbar}$    \\
      \hline
      $8.00$              & $4.2(1)$  & $127(1)$    \\
      $8.45, 16^3\,32$    & $3.5(2)$  & $118(3)$    \\
      $8.45, 24^3\,48$    & $4.1(1)$  & $119(1)$    \\
    \end{tabular}
  \caption{
    \label{tab:mq}
    Light and strange quark masses}
\end{minipage}
\end{table}
Note, that the lattice artefacts (represented by the slope) scale with
$a^2$, as expected for (the automatically ${\cal{O}}(a)$ improved) overlap
fermions.  The resulting quark masses are given in~Table~\ref{tab:mq}.
The light quark mass scales very well, but we see a volume dependence. For
the strange quark mass we see some dependence on $a$, but not on the volume.

\section{Vector meson and nucleon masses}
\label{sec:scaling}

\begin{figure}[tbh]
\vspace{-8mm}  \begin{minipage}[t]{0.45\linewidth}
    \epsfig{file=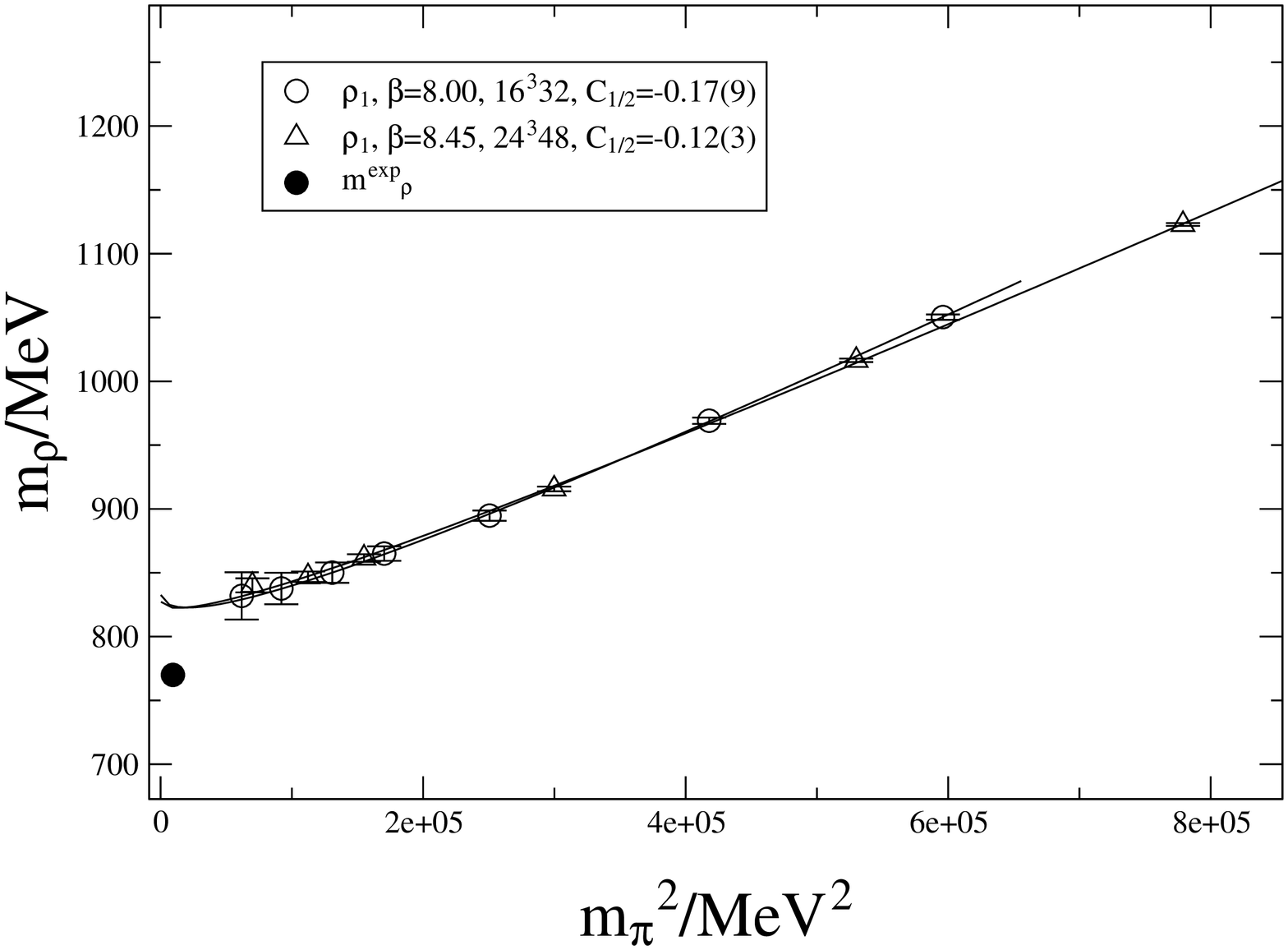,width=5.5cm, angle=270, clip=}
    \caption{
      \label{fig:mrho}
      Quenched chiral fits to the vector meson.}
  \end{minipage}
  \hspace{0.05\linewidth}
  \begin{minipage}[t]{0.45\linewidth}
    \epsfig{file=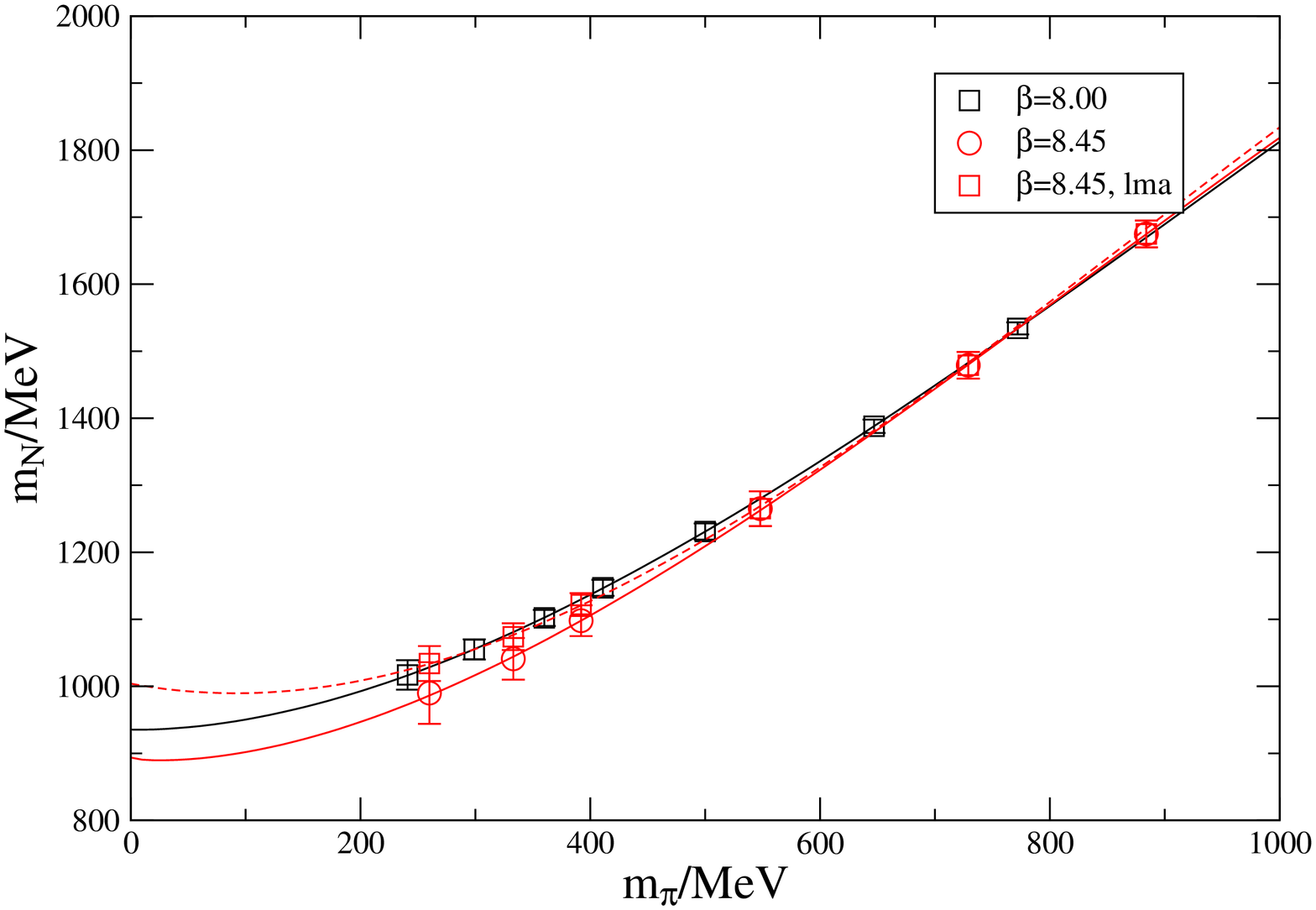,width=5.5cm,angle=270, clip=}
    \caption{
      \label{fig:mN}
      The same for the nucleon. The dashed line is a fit to low mode averaged
      data.}
  \end{minipage}
\end{figure}

In this section we extract vector meson and nucleon masses from our data and
compare their chiral behaviour to predictions of quenched chiral perturbation theory.
The vector meson mass (Fig.~\ref{fig:mrho}) confirms the prediction of a negative value of $C_{1/2}$
in a fit according to
$m_\rho(m_\pi)=m_\rho+C_{1/2}m_\pi+C_1m_\pi^2+C_{3/2}m_\pi^3$.
The prediction for the nucleon has the same structure, but it is harder to
confirm the negative sign of $C_{1/2}$ for $m_N$, since the errors make the
results compatible with zero unless one uses the low mode averaging technique.
In this case we find a value of $C_{1/2}=-0.32(18)$ at $\beta=8.45$, $24^3\,
48$, which is in good agreement with the chiral perturbation theory prediction
$C_{1/2}=3/2\pi(D-3F)^2 \delta=-0.25(5)$\footnote{D and F are axial current
  matrix elements, D=0.81(3), F=0.47(4)~\cite{Jaffe:1989jz}} (Fig.~\ref{fig:mN}).
\section{Summary}
\label{sec:summary}

In a quenched overlap simulation we have computed the light and strange quark masses
$m_\ell^\msbar(2\gev)=4.1(1)\mev$ and $m_s^\msbar(2\gev)=119(1)\mev$. We
demonstrated the existence of quenched chiral logs and the expected chiral
behaviour of $f_\pi$, $m_\rho$, and $m_N$.  

\section{Acknowledgements}
\label{sec:acknowledgements}

The numerical calculations were performed at the IBM pSeries 690 computers at
HLRN and NIC J\"ulich, and on the PC farm at DESY Zeuthen.We thank these
institutions for their support. Part of this work is supported by the DFG
under contract FOR465. 
\bibliography{bib}

\providecommand{\href}[2]{#2}\begingroup\raggedright\begin{thebibliography}{1}

\bibitem{streuer:_nucleon}
{\bf QCDSF} Collaboration, T.~Streuer {\em et.~al.}, {\it Nucleon structure
  from quenched overlap fermions},  {\em PoS(LAT2005)363 (these proceedings)}.

\bibitem{Giusti:2002sm}
L.~Giusti, C.~Hoelbling, M.~L{\"u}scher and H.~Wittig, {\it Numerical
  techniques for lattice QCD in the {$\varepsilon$}- regime},  {\em Comput.
  Phys. Commun.} {\bf 153} (2003) 31--51
  [\href{http://arXiv.org/abs/hep-lat/0212012}{{\tt hep-lat/0212012}}].

\bibitem{Luscher:1985xn}
M.~L{\"u}scher and P.~Weisz, {\it On-shell improved lattice gauge theories},
  {\em Commun. Math. Phys.} {\bf 97} (1985) 59.

\bibitem{Gattringer:2001jf}
C.~Gattringer, R.~Hoffmann and S.~Schaefer, {\it Setting the scale for the
  L{\"u}scher-Weisz action},  {\em Phys. Rev.} {\bf D65} (2002) 094503
  [\href{http://arXiv.org/abs/hep-lat/0112024}{{\tt hep-lat/0112024}}].

\bibitem{Horsley:2004mx}
{\bf QCDSF} Collaboration, R.~Horsley, H.~Perlt, P.~E.~L. Rakow, G.~Schierholz
  and A.~Schiller, {\it One-loop renormalisation of quark bilinears for overlap
  fermions with improved gauge actions},  {\em Nucl. Phys.} {\bf B693} (2004)
  3--35 [\href{http://arXiv.org/abs/hep-lat/0404007}{{\tt hep-lat/0404007}}].
  Erratum-ibid. {\bf B713} (2005) 601.

\bibitem{Martinelli:1994ty}
G.~Martinelli, C.~Pittori, C.~T. Sachrajda, M.~Testa and A.~Vladikas, {\it A
  General method for nonperturbative renormalization of lattice operators},
  {\em Nucl. Phys.} {\bf B445} (1995) 81--108
  [\href{http://arXiv.org/abs/hep-lat/9411010}{{\tt hep-lat/9411010}}].

\bibitem{Gockeler:1998ye}
M.~G{\"o}ckeler {\em et.~al.}, {\it Nonperturbative renormalisation of
  composite operators in lattice QCD},  {\em Nucl. Phys.} {\bf B544} (1999)
  699--733 [\href{http://arXiv.org/abs/hep-lat/9807044}{{\tt
  hep-lat/9807044}}].

\bibitem{Gracey:2003yr}
J.~A. Gracey, {\it Three loop anomalous dimension of non-singlet quark currents
  in the $RI'$ scheme},  {\em Nucl. Phys.} {\bf B662} (2003) 247--278
  [\href{http://arXiv.org/abs/hep-ph/0304113}{{\tt hep-ph/0304113}}].

\bibitem{Jaffe:1989jz}
R.~L. Jaffe and A.~Manohar, {\it The g(1) problem: fact and fantasy on the spin
  of the proton},  {\em Nucl. Phys.} {\bf B337} (1990) 509--546.

\end{thebibliography}\endgroup
\end{document}